\documentclass[reprint,footinbib,amsmath,amssymb,aps,prl,floatfix,twocolumn,superscriptaddress]{revtex4-2}

\usepackage{physics}
\usepackage{printlen} 
\usepackage{amsthm}
\usepackage{dsfont}
\usepackage{bbm}
\usepackage{mathtools}

\usepackage[normalem]{ulem}
\usepackage{makecell}

\usepackage{stmaryrd}

\usepackage{graphicx}
\graphicspath{{images/}}
\usepackage{wrapfig}

\usepackage{dcolumn}
\usepackage{bm}
\usepackage[hidelinks]{hyperref}
\usepackage{tikz}
\usepackage{dsfont}
\usetikzlibrary{quantikz}

\usepackage{diagbox}

\usepackage{lipsum} 

\usepackage[labelformat=simple]{subcaption}

\usepackage{ragged2e}
\DeclareCaptionJustification{justified}{\justifying}
\usepackage{printlen}

\usepackage{etoolbox}

\usepackage[innercaption]{sidecap}
\sidecaptionvpos{figure}{c}

\usepackage{dsfont}

\makeatletter
\def\maketitle{
\@author@finish
\title@column\titleblock@produce
\suppressfloats[t]}
\makeatother

\newcounter{SMsections}
\renewcommand{\theSMsections}{\Roman{SMsections}}
\DeclareRobustCommand{\SMsec}[2]{%
    \begin{center}
        \medskip
        \refstepcounter{SMsections}%
        \addcontentsline{toc}{section}{\theSMsections.\space#1}
        \textbf{\theSMsections.\quad\label{#2} \uppercase{#1}}
    \end{center}
}

\newcommand{\n}[1]{\mathrm{#1}}
\newcommand{\cl}[1]{\mathcal{#1}}
\newcommand{\bb}[1]{\mathbb{#1}}
\newcommand{\msf}[1]{\mathsf{#1}}

\newcommand{\id}{\mathds{1}}

\newcommand{\be}{\begin{equation}} 
\newcommand{\ee}{\end{equation}}

\newcommand{\Gerr}{\ensuremath{\Gamma_\n{e}}}
\newcommand{\Gqec}{\ensuremath{\Gamma_\n{c}}}

\newcommand{\ketL}[1]{\ket{#1_\n{L}}}

\newcommand{\overbar}[1]{\mkern 1.5mu\overline{\mkern-1.5mu#1\mkern-1.5mu}\mkern 1.5mu}

\newtoggle{arXiv} 
\toggletrue{arXiv}      

\begin{document}
\preprint{APS/123-QED}

\title{Scalable dissipative quantum error correction for qubit codes}

\author{Ivan Rojkov}
\email{irojkov@phys.ethz.ch}
\thanks{Present address: Yale Quantum Institute, Yale University, New Haven, Connecticut 06520, USA}
\affiliation{Institute for Quantum Electronics, ETH Z\"{u}rich, Otto-Stern-Weg 1, 8093 Z\"{u}rich, Switzerland}
\affiliation{Quantum Center, ETH Zürich, 8093 Zürich, Switzerland}
\author{Elias Zapusek}
\affiliation{Institute for Quantum Electronics, ETH Z\"{u}rich, Otto-Stern-Weg 1, 8093 Z\"{u}rich, Switzerland}
\affiliation{Quantum Center, ETH Zürich, 8093 Zürich, Switzerland}
\author{Florentin Reiter}
\affiliation{Institute for Quantum Electronics, ETH Z\"{u}rich, Otto-Stern-Weg 1, 8093 Z\"{u}rich, Switzerland}
\affiliation{Quantum Center, ETH Zürich, 8093 Zürich, Switzerland}
\affiliation{Fraunhofer Institute for Applied Solid State Physics IAF, Tullastr. 72, 79108 Freiburg, Germany}

\date{\today}


\begin{abstract}
    Dissipative quantum error correction (QEC) autonomously protects quantum information using engineered dissipation and offers a promising alternative to error correction via measurement and feedback. However, scalability remains a challenge, as correcting high-weight errors typically requires increasing dissipation rates and exponentially many correction operators. Here, we present a scalable dissipative QEC protocol for discrete-variable codes, correcting multi-qubit errors via a trickle-down mechanism that sequentially reduces errors weight. Our construction exploits redundancy in the Knill--Laflamme conditions to design correction operators that act on multiple error subspaces simultaneously, thereby reducing the overhead from exponential to polynomial in the number of required operators. We illustrate our approach with repetition codes under biased noise, showing a fourfold improvement in the exponential suppression factor at realistic physical error rates.
\end{abstract}

\maketitle


\noindent
\textit{Introduction.---}
Quantum error correction (QEC) is an essential ingredient for fault tolerant quantum information processing~\cite{aharonov_fault_1999}. Standard QEC techniques rely on classical control theory, requiring auxiliary quantum systems, measurements of error syndromes, and feedback-conditioned operations to correct errors in the target system. Although recent experiments have demonstrated the feasibility of these QEC schemes~\cite{ryan-anderson_realization_2021,krinner_realizing_2022,bluvstein_logical_2024} and achieved logical lifetimes greater than of best physical component in the system (known as break-even point)~\cite{acharya_quantum_2024,paetznick_demonstration_2024}, they often rely on correction in post-processing or show limited advantage over uncorrected quantum systems. The main limitations arise from errors introduced during syndrome measurements~\cite{ofek_extending_2016,hu_quantum_2019,fluhmann_encoding_2019,rosenblum_fault-tolerant_2018} and the time delays caused by feedback loops~\cite{reilly_challenges_2019,campagne-ibarcq_quantum_2020,terhal_quantum_2015}.

Dissipative, or autonomous, QEC~\cite{paz_continuous_1998,ahn_continuous_2002,sarovar_continuous_2005,oreshkov_chapter8_2013,lebreuilly_autonomous_2021} represents an alternative paradigm in which measurements and standard recovery operations are replaced by coherent dynamics and reset channels. The most prominent experimental realizations have been achieved with bosonic codes, which encode quantum information in systems with extended Hilbert spaces such as harmonic oscillators~\cite{campagne-ibarcq_quantum_2020,de_neeve_error_2022,leghtas_confining_2015,touzard_coherent_2018,grimm_stabilization_2020,lescanne_exponential_2020,gertler_protecting_2021,rousseau_enhancing_2025}. Extending dissipative QEC to qubit codes, the dominant class of codes that rely on a redundant use of two-level systems, has proven more challenging. Although many theoretical proposals exist~\cite{kerckhoff_designing_2010,pastawski_quantum_2011,cohen_dissipation_2014,ippoliti_perturbative_2015,kapit_hardware-efficient_2016,hsu_method_2016,reiter_dissipative_2017,lihm_implementation_2018,heusen_measurement-free_2024}, only a few small scale experimental demonstrations have been achieved to date~\cite{livingston_experimental_2022,li_autonomous_2024}. The main challenge is their ability to correct high-weight errors, i.e. faults involving many elementary (typically single-qubit) errors separating a state from the codespace. Recent studies~\cite{lebreuilly_autonomous_2021,shtanko_bounds_2025} suggest that dissipative QEC can, in principle, yield exponential suppression of logical error rates with increasing code size, but only at the cost of engineering an exponential number of correction operators applied at rates that scale polynomially with system size.

In this Letter, we show that this inefficiency originates from designing distinct correction operators for each error subspace, whose number grows exponentially with system size. To overcome this, we introduce a scalable dissipative QEC construction for qubit codes that operates in a \textit{trickle-down} fashion, driving the system through successive lower-weight error subspaces until it reaches the codespace. It is enabled by the observation that the Knill--Laflamme conditions~\cite{knill_theory_1997}, usually formulated between the codespace and error subspaces, also hold between different error subspaces. When two such pairs satisfy these conditions identically, a single operator can correct both. This allows correction operators to simultaneously reduce the weight of many error subspaces, leading ultimately to an exponential reduction in overhead: the required number of operators scales only polynomially with code size. We illustrate the approach using repetition codes under biased noise~\cite{aliferis_fault-tolerant_2008} with a scheme realizable in a trapped ion architecture. We show the exponential improvement in the correction performance compared to standard dissipative QEC methods and relate our approach to current developments in QEC.

\begin{figure}[b]
    \includegraphics{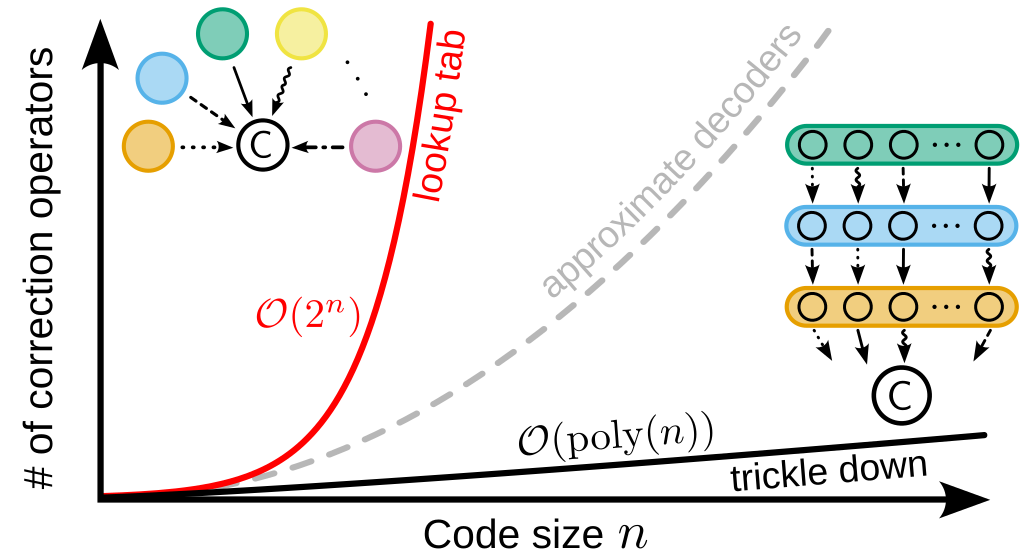}
    \caption{\textit{Number~of~correction~operators~vs.~code~size.} Schematic comparison of the number of jump operators required to correct an $n$-qubit code $\msf{C}$ using lookup-table (red), trickle-down (black), and approximate decoder (dashed) schemes. Lookup tables assign one operator per error subspace, scaling as $\cl{O}(2^n)$. The trickle-down approach groups subspaces by error weight and use operators that reduce the weight of many subspaces simultaneously, yielding polynomial scaling. Approximate decoders offer an intermediate solution.} 
    \label{fig1}
\end{figure}


\medskip\noindent
\textit{Setting.---}
We consider the problem of protecting a logical subspace~$\msf{C}$ of a Hilbert space $\msf{H}$ from incoherent noise that disturbs the system. We model the dynamics responsible for the errors and their correction using a Lindblad equation $\dot{\rho} = (\cl{L}_\n{e} + \cl{L}_\n{c})(\rho)$. Both Liouvillians describe purely incoherent processes ${\cl{L}_\n{e}(\rho) := \sum_i \cl{D}[L_{\n{e},i}](\rho)}$ with dissipators $\cl{D}[L](\rho) = L \rho L^\dagger - \frac{1}{2}\left(L^\dagger L \rho + \rho L^\dagger L \right)$ (similarly for~$\cl{L}_\n{c}$). The jump operators $\{L_{\n{e},i}\}$ describe errors caused by unwanted system-environment couplings, while $\{L_{\n{c},i}\}$ represent the corrections. Dissipative QEC aims to engineer the latter so that the logical subspace $\msf{C}$ is stabilized under both Liouvillians.

The form of the correction operators is determined by the encoding of the logical subspace $\msf{C}$ into the physical Hilbert space $\msf{H}$. Here, we consider a system of $n$ qubits, i.e. $\msf{H}=\mathbb{C}^{2^n}$, and assume $\sf{C}$ to be $k$-dimensional spanned by the logical states $\{\ketL{\mu}\}_{\mu=0}^{k-1}$. The code choice is not unique and depends on the error set $\bb{E}=\{E_i\}$ against which we want it to be robust. Each $E_i$ corresponds to the accumulated error operator defined by a product of operators drawn from $\bb{L}=\{L_{\n{e},i}\}\!\bigcup\{L_{\n{e},i}^\dagger L_{\n{e},i}\}$. To correct these errors, $\msf{C}$ must satisfy the Knill--Laflamme condition $P_\msf{C}E_j^\dagger E_i P_\msf{C}=p_i\delta_{ij}P_\msf{C}$ $\forall i,j$ with $p_i$ quantifying the detection probability of $E_i$ and $P_\msf{C}$ the projector on the logical subspace~\cite{knill_theory_1997}. We assume, without loss of generality, its diagonal form~\footnote{See Supplemental Material, which includes Refs.~\cite{higgott_sparse_2025,edmonds_maximum_1965,edmonds_paths_1965,dennis_topological_2002,fowler_towards_2012,iolius_decoding_2024,jones_improved_2024,ott_decision-tree_2025,beni_tesseract_2025,pantaleoni_modular_2020,mirrahimi_dynamically_2014,gilles_generation_1994,chamberland_building_2022,stabilization_rojkov_2024,gottesman_encoding_2001,royer_stabilization_2020,sivak_real-time_2023,shaw_stabilizer_2024,de_neeve_modular_2025,siegele_robust_2023}, for more detailed derivations and further discussions of the results presented in the main text.}. The recovery operation for the code $\msf{C}$ can then be written as a completely positive trace preserving map $\cl{R}(\rho) = \sum_{i} U_i^\dagger P_i \, \rho \, P_i U_i + P_\msf{E} \, \rho \, P_\msf{E}$ where $P_i=U_i P_\msf{C} U_i^\dagger \equiv E_i P_\msf{C} U_i^\dagger/\sqrt{p_i}$ and ${P_\msf{E}=\id-\sum_i P_i }$ are the projectors onto the error subspace associated with $E_i$ and onto the subspace of errors violating the correctability condition, respectively. The unitaries $\{U_i\}$ are obtained through a polar decomposition of $E_i$ and, following the Knill--Laflamme condition, are mutually orthogonal $P_\msf{C}U_j^\dagger U_i P_\msf{C}=\delta_{ij}P_\msf{C}$ $\forall i,j$~\cite{leung_approximate_1997}. Measurement-feedback QEC implements $\cl{R}$ by coupling the system to an auxiliary one whose measurement projects the state onto an error subspace. The outcomes are then processed by a classical decoder to apply the corresponding recovery operation $U_i^\dagger$, returning the state to the codespace. 

In the dissipative QEC setting, these processes are performed by $\cl{L}_\n{c}$. A naive solution for the correction jump operators is to consider a set $\{L_{\n{c},j}\}$ such that $\cl{L}_\n{c}(\rho)=\Gqec(\cl{R}(\rho) - \rho)$ with $\Gqec$ being the correction rate. From the construction of the correction map $\cl{R}$ and assuming $P_\msf{E}\equiv0$, we can see that the jump operators
\be \label{eq:jump_lookup_table}
    L_{\n{c},i}=\sqrt{\Gqec}\,U_i^\dagger\,P_i
\ee
realize the desired Liouvillian~\footnote{If $P_\msf{E}\neq0$ (e.g., in nonadditive quantum code), additional jump operators are necessary to prevent the system to get stuck in the uncorrected subspace~\cite{lihm_implementation_2018,lebreuilly_autonomous_2021}.}. Concretely, a $\llbracket n,k,d \rrbracket$ stabilizer code, for example, encoding $k$ logical qubits in an $n$-qubit system and robust up to $d$ errors, lives in a $2^k$-dimensional subspace of $\msf{H}$ that is invariant under the group generated by $n-k$ independent operators $S_i$. Using these generators, we can construct the set of projectors $\{P_i\}_{i=1}^{N_\n{sub}}$ onto $N_\n{sub}=2^{n-k}-1$ nontrivial and orthogonal error subspaces. The set $\{L_{\n{c},i}\}$ results then in $N_\n{sub}=\cl{O}(2^n)$ jump operators~\cite{Note2}. Implementation proposals for dissipative QEC have largely focused on this construction~\cite{paz_continuous_1998,ahn_continuous_2002,sarovar_continuous_2005,oreshkov_chapter8_2013,ippoliti_perturbative_2015,hsu_method_2016,reiter_dissipative_2017,lihm_implementation_2018}.

This construction has two fundamental limitations: (1) its experimental implementation has an exponential operational cost as it is equivalent to a lookup-table decoder~\cite{Note1}; (2) the correction rate $\Gqec$ must scale as $\cl{O}(\n{poly}(n))$ to maintain a constant ratio $p_\n{e}/p_\n{c}$ between the probabilities of occurrence of an error and a correction jump (see End Matter)~\cite{pastawski_quantum_2011,shtanko_bounds_2025}. We will refer to this approach as the \textit{lookup-table} dissipative QEC.

\begin{figure*}[ht!]
    {\phantomsubcaption\label{fig:threshold}}
    {\phantomsubcaption\label{fig:exp_supp_factor}}
    {\phantomsubcaption\label{fig:num_of_qubits}}
    \includegraphics{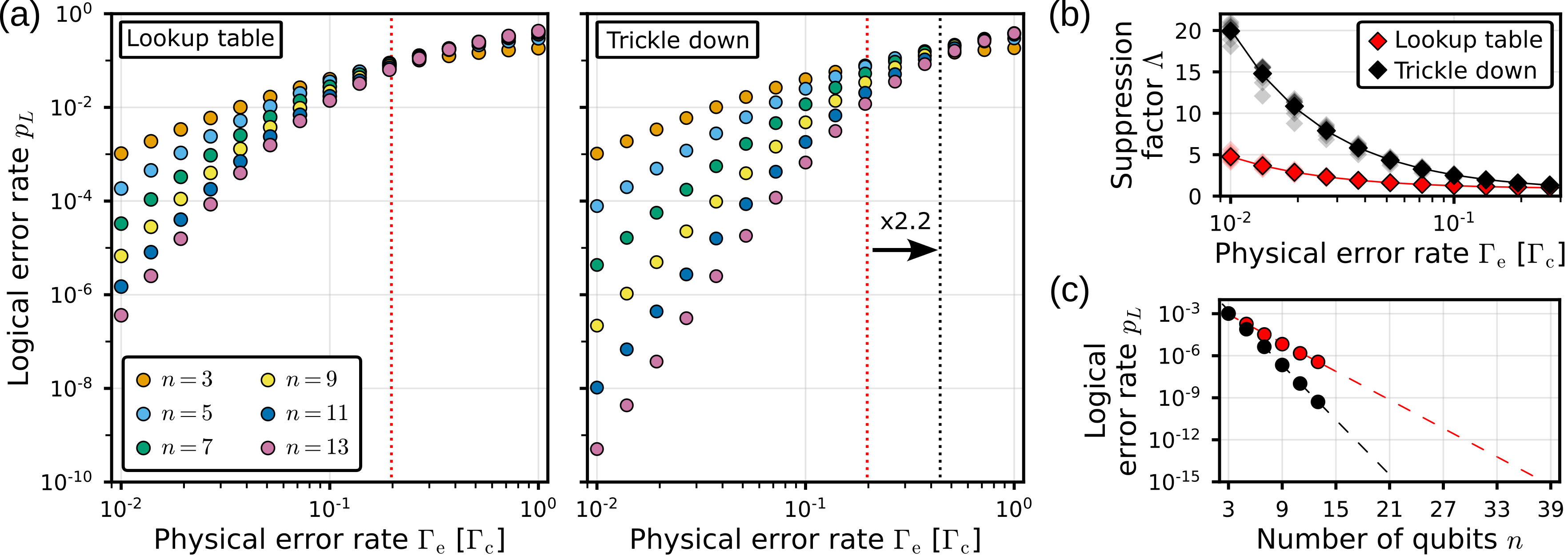}
    \caption{\textit{Trickle-down~vs.~lookup-table~QEC.} (a) Logical error rate $p_L$ as a function of the physical error rate $\Gerr$ for lookup-table (left) and trickle-down (right) dissipative QEC of $n$-qubit repetition codes under bit-flip noise. Vertical lines indicate the threshold values $\sim0.2$ and $\sim0.44$ for both solutions, respectively. (b) Exponential suppression factor $\Lambda=\Gqec/\Gerr^*$ as a function of $\Gerr$ for the lookup-table (red) and trickle-down (black) approaches. Data points correspond to $\big(p_{L,i}(\Gerr)/p_{L,i+k}(\Gerr)\big)^{1/k}$, with $p_{L,i}(\Gerr)$ the logical error rate for an $i$-qubit code at fixed $\Gerr$. Shaded markers show individual data points; solid markers show their averages. Solid lines are fits to inverse scaling, yielding $\Gerr^* \approx 0.04$ and $\Gerr^* \approx 0.2$ for lookup-table and trickle-down schemes, respectively. (c) Code sizes required to achieve specific $p_L$ values at a fixed physical error rate $\Gerr = 10^{-2} \Gqec$. Dashed lines show extrapolations based on $\Gerr^*$. }
    \label{fig3}
\end{figure*}

\medskip\noindent
\textit{Trickle-down construction.---}
While the number of error subspace $N_\n{sub}$ scales exponentially with the system size, the number of elementary error jumps $L_{\n{e},i}$ often increases only polynomialy with $n$. We thus partition $\bb{E}$ into $i$th-order error subsets $\bb{E}_i$ such that ${\bb{E}_0=\{\id\}}$, ${\bb{E}_1=\bb{L}}$, $\bb{E}_2=\{O_iO_j\,|\,{O_k\in\bb{L}},\,\, {i \neq j}\}$, and $\bb{E}=\bigcup_i \bb{E}_i$, and proceed identically with error subspaces and their projectors, i.e. ${\bb{P}_0=\{P_\msf{C}\}}$, ${\bb{P}_1=\{P_i\text{ associated to }E_i\in\bb{E}_1\}}$, etc. Similar to the Knill--Laflamme condition setting the relationship between the codespace and individual error subspaces, we derive a condition on the relationship between the error subspaces themselves: 

A $q$th-order error subspace with $P_i^{(q)}\in\bb{P}_q$ associated to $E_i^{(q)}\in\bb{E}_q$ is transformed to a $(q+p)$th-order error subspace under the action of $p$th-order errors. Assuming that $\msf{C}$ is capable of correcting errors up to order ${\ell \geq q+p}$, we get the following condition  
\be \label{eq:KL_error_subspaces}
\begin{split}
    &P_i^{(q)}E_j^{(p)\dagger}E_k^{(p)}P_i^{(q)}  \\
    &\propto U_i^{(q)} \Big[P_\msf{C} E_i^{(q)\dagger}E_j^{(p)\dagger} E_k^{(p)}E_i^{(q)} P_\msf{C}\Big] U_i^{(q)\dagger} \\
    &\propto U_i^{(q)} \delta_{jk} P_\msf{C} U_i^{(q)\dagger} =\, \delta_{jk} P_i^{(q)}\,.
\end{split}
\ee  
where we use the definition of the projectors $P_i$, the recovery unitary $U_i^{(q)}$ from the polar decomposition of $E_i^{(q)}$, and the Knill--Laflamme condition for the $(q+p)$th-order errors. Consequently, the codespace $\msf{C}$ and a $(q+p)$th-order error subspace can both be recovered with the same unitary $U_j^{(p)\dagger}$,
\be \label{eq:jump_2_subspaces}
    L_{\n{c},i}^{(p)} = \sqrt{\Gqec} \, U_i^{(p)\dagger} \left( P_i^{(p)} + P_i^{(q+p)} \right)\,.
\ee
In this manner, we can design a set of jump operators that correct errors in a trickle-down fashion. The minimal set has a size bounded by the cardinality of $\bb{L}$, i.e. the number of elementary errors, denoted $\abs{\bb{L}}$, and is composed of the correction operators
\be \label{eq:jump_optimal}
    L_{\n{c},i}^{(1)} = \sqrt{\Gqec} \, U_i^{(1)\dagger} \sum_{j=1}^{\ell} \sum_{k=1}^{K_j} P_{i,k}^{(j)} \quad \forall i \in \{1,\ldots,\abs{\bb{L}}\}
\ee
where $U_i^{(1)\dagger}$ are the recovery unitaries associated to the error ${E_i^{(1)}\!\in{\bb{E}_1\equiv\bb{L}}}$ and ${P_{i,k}^{(j)}\in\bb{P}_j}$ are the projectors onto $j$th-order error subspaces $\msf{P}_i^{(j)}$ related to a $(j-1)$th-order subspaces through the error process $E_i^{(1)}$, i.e. ${P_{i,k}^{(j)}=U_i^{(1)}P_k^{(j-1)}U_i^{(1)\dagger}}$ for some index $k$. This index runs from $1$ to ${K_j=\abs{\bb{P}_{j-1}}-1}$. The total number of projectors $P_{i,k}^{(j)}$ in each jump operator $i$ is given by the combinatorial $N_\n{proj} = \sum_{j=1}^{\ell} \binom{\abs{\bb{L}}-1}{j-1}$ which scales exponentially with $\ell/\abs{\bb{L}}$ (see End Matter).

The ratio of the probabilities that an error versus a correction operator occurs is given by 
\be \label{eq:proba_err_corr_optimal}
    \frac{p_\n{e}}{p_\n{c}} = \frac{N_e\Gerr}{N_\n{proj}\,\Gqec}\,.
\ee
Unlike in the lookup-table case (see End Matter), this ratio explicitly depends on $\msf{C}$, through the maximum correctable error order $\ell$ and the number of projectors $N_\n{proj}$. Consequently, two codes correcting errors up to the same order $\ell$ but with different numbers of physical qubits (e.g., Shor and Steane codes) would require different numbers of jump operators and exhibit different correction performance. This contrasts sharply with the lookup-table scheme, where performance depends only on the correction rate $\Gqec$.

\medskip\noindent
\textit{Repetition code.---}
We illustrate these results using $\llbracket n,1,n \rrbracket$ repetition codes, encoding a single logical qubit in $n$ physical ones and protecting against single qubit bit-flip errors $X_i$ up to order $\ell=(n-1)/2$. The codespace $\msf{C}$ is spanned by the states $\{\ket{0_\n{L}}=\ket{0}^{\otimes n},\ket{1_\n{L}}=\ket{1}^{\otimes n}\}$. The projectors onto the error subspaces can be constructed using the generators of the code's stabilizer group, $\{Z_1 Z_2,Z_2 Z_3, ..., Z_{n-1} Z_n\}$. We use these to build the correction operators using either the lookup-table approach in Eq.~\eqref{eq:jump_lookup_table} or the trickle-down construction in Eq.~\eqref{eq:jump_optimal}. Ignoring their potentially nonlocal nature, the former requires the engineering of $n$ jump operators, whereas the lookup-table solution demands the implementation of ${2^{n-1}-1}$.

The two constructions also differ in their error correction performance. We study it by simulating the dynamics of repetition codes initialized in the state $\ket{\psi_\n{L}}\propto\ket{0_\n{L}}+i\ket{1_\n{L}}$, subjected to bit-flip errors ${L_{\n{e},i}=\sqrt{\Gerr}\,X_i}$ and correction jumps given by either Eq.~\eqref{eq:jump_lookup_table} or Eq.~\eqref{eq:jump_optimal}. After an evolution time $t<3/\Gerr$ we decode the system state by majority voting and compute the logical fidelity with the initial state~\cite{Note1}. The long timescale trend of the infidelity curves can be described by the function ${F(t)=0.5\left[1+\exp(\epsilon\,(t-\tau))\right]}$ where $\epsilon$ represents the rate of the decay and $\tau$ the time at which this long timescale regime starts. This function is derived from a combinatorial counting of the total number of errors that occurred in the system~\cite{obrien_density-matrix_2017}. Ideally both parameters are low, therefore the dimensionless quantity $p_L=\epsilon\,\tau$ representing the probability to misidentify the logical state is a good measure of the error correction performance. When $\tau \propto 1/\Gerr$, $p_L$ becomes the logical error rate commonly used in measurement-based QEC where it represents the frequency with which the decoder fails. We evaluate the error correction performance of the two approaches by fitting this function to the infidelity curves using the Levenberg--Marquardt algorithm that we bias using $t^{\,\ell+1}$ to avoid overfitting the dynamics at short times~\cite{obrien_density-matrix_2017}. We do this for various code sizes $n\in\{3,5,7,9,11,13\}$ and error rates $\Gerr\in\big[10^{-2},10^{0}\big]\times\Gqec$. Fig.~\ref{fig:threshold} illustrate $p_L$ as a function of $\Gerr$ for both solutions. Both situations show the presence of a threshold below which increasing the code size, exponentially decreases the logical error rate. Yet, the threshold of the trickle-down solution is $\sim2.2$ larger than the lookup-table one (see details below). Moreover, in the sub-threshold regime, we observe that the trickle-down solution outperforms the lookup-table one for all ${n>3}$~\footnote{For $n=3$, both solutions are identical.}. In this regime, the logical error rate follows $p_L \propto (\Gerr/\Gerr^*)^{\ell+1}$ with $\Gerr^*$ a constant related to the threshold value. A figure of merit is the exponential error suppression factor $\Lambda=\Gerr^*/\Gerr$ which can be obtained from the ratio of $p_L$ for different $n$~\cite{kelly_state_2015,chen_exponential_2021,acharya_suppressing_2023}. Specifically, we calculate $\big(p_{L,i}(\Gerr)/p_{L,i+k}((\Gerr))\big)^{1/k}$ for different $k$ with $p_{L,i}(\Gerr)$ being the logical error rate for an $i$-qubit code and an error rate $\Gerr$. These quantities are then averaged out to estimate the exponential suppression factor for a given $\Gerr$. We plot $\Lambda$ as a function of the error rate in Fig.~\ref{fig:exp_supp_factor} for both solutions and show that the trickle down one allows to increase $\Lambda$ by a factor of $\sim4$. By fitting an inverse proportionality function $f(\Gerr)= A/\Gerr + B $~\footnote{The offset $B$ is here to capture the behavior of $\Lambda$ in the high error rate regime.} to these data, we estimate $\Gerr^*=0.04$ for the lookup table solution and $\Gerr^*=0.2$. Coincidentally, the positions of the thresholds (dotted lines in Fig.~\ref{fig:threshold}) are located at the square root of $\Gerr^*$. As a final performance metric, we use $\Gerr^*$ to extrapolate the number of qubits $n$ required to reach a logical error rate $p_L=10^{-15}$ given a physical error rate $\Gerr=10^{-2}$. The results are presented in Fig.~\ref{fig:num_of_qubits} where the dashed lines follow $p_L \propto (\Gerr/\Gerr^*)^{\ell+1}$ with proportionality constants $0.02$ and $0.7$ for the lookup-table and trickle-down solutions, respectively. The target $p_L$ is achieved with $\sim21$ qubits using the trickle-down solution, compared to $\sim37$ qubits for the lookup-table approach.

\medskip\noindent
\textit{Trapped-ion scheme.---}
To implement a trickle-down correction of a repetition code, we extend the reservoir-engineering scheme of Ref.~\cite{reiter_dissipative_2017} and show that it naturally realizes the jump operators in Eq.~\eqref{eq:jump_optimal}. We consider a system of $n$ ions each possessing four addressable internal energy levels, $\{\ket{0},\ket{1},\ket{e},\ket{f}\}$, coupled to three motional modes $a$, $b$ and $c$ of the ion string. The computational subspace is spanned by $\{\ket{0},\ket{1}\}^{\otimes n}$, while the additional levels and modes are used for the reservoir engineering of the correction jump operators, i.e. the engineering of a Hamiltonian $H$ and a tunable nonunital (i.e. entropy removing) dissipation channel (e.g., optical pumping or sympathetic cooling~\cite{van_mourik_experimental_2024}). In our case, $H$ consists of three laser fields: $H_1 = \sum_{i=1}^{n}  \Delta \left( \dyad{e}{e}_i + \dyad{f}{f}_i \right) + \frac{\Omega}{2} \left( \dyad{e}{0}_i + \dyad{f}{1}_i \right) + \n{H.c.}$; $H_2 = \delta \left( a^\dagger a + b^\dagger b \right) + G \sum_{i=1}^{n} a^\dagger \dyad{1}{e}_i + b^\dagger \dyad{0}{f}_i + \n{H.c.}$; and $H_{3,i} = g\,c_i^{\dagger} \left( \dyad{0}{e}_i + \dyad{1}{f}_i \right) + \n{H.c.}$ The first two correspond to global lasers driving all ions simultaneously, while the last one is a single-ion addressing field used to remove entropy from the system through jump operators $L_{i}=\sqrt{\kappa} \, c_i$ with a tunable rate $\kappa$. In the Supplemental Material, we show that the effective dynamics of the system is dissipative with a jump operator
\be \label{eq:eff_jump_op_proj}
\begin{split}
    L_{\n{eff},i}
    =\sum_{j=1}^\ell &\sqrt{\kappa_\n{eff}(j)} \big(\dyad{0}{1}_i P_{1}^{(j)} + \dyad{1}{0}_i P_{0}^{(j)}\big) \\
    +&\sqrt{\kappa_\n{eff}(n-j)} \big(\dyad{1}{0}_i P_{1}^{(j)} + \dyad{0}{1}_i P_{0}^{(j)}\big)\,.
\end{split}
\ee
where $P_{\psi}^{(j)}$ indicate projectors onto all computational state that are $j$ bit flips away from the logical state $\ketL{\psi}$. The first parenthesis in $L_{\n{eff},i}$ displays the \textit{desired} error correction operators, while the second one bares the \textit{undesired} processes increasing the error weight instead of decreasing it. The rate of these two processes follow a Lorentzian function (see Supplemental Material)
\be \label{eq:eff_jump_op_rate}
    \kappa_\n{eff}(x) = \frac{\Omega^2}{\kappa_\n{eng}} 
    \left[ \left(\frac{2 G}{\kappa_\n{eng}\delta}\right)^{\!2} \left(x - \frac{\delta\Delta}{G} \right)^{\!2} + 1\, \right]^{-1}
\ee
with ${\kappa_\n{eng}\propto g^{2}/\kappa}$ being the rate of the engineered dissipation. Choosing $x_0=\delta\Delta/G$ to be an integer, the process will correct all error states of weight $x_0$ with erroneous qubit $i$ resonantly at a rate $\Omega^2/\kappa_\n{eng}$. All the other processes will be off resonant. Although Eq.~\eqref{eq:eff_jump_op_proj} is qualitatively different than the trickle-down solution in Eq.~\eqref{eq:jump_optimal}, they can be made equivalent  by modifying $H_2$ to be a multi-tone drive (see End Matter). Finally, a full correction would require $n$ single-ion addressing laser beams with $n$ auxiliary motional mode $c_i$, i.e. a Hamiltonian $H_{3,i}$ for every ion. This would engineer a dissipative process with $n$ jump operators $\{L_{\n{eff},i}\}_{i=1}^n$. Although challenging, this scheme demonstrates that the trickle-down dissipative QEC can be physically implemented. In fact, the data presented in Fig.~\ref{fig3} were simulated using $\{L_{\n{eff},i}\}_{i=1}^n$ with experimentally achievable parameters values, $\delta=10\kappa_\n{eng}$, $\Omega=3\kappa_\n{eng}$, $\kappa_\n{eng}=10~\n{kHz}$.

\medskip\noindent
\textit{Discussion and conclusion.---}
In summary, we introduced a scalable construction of jump operators for dissipative QEC based on a trickle-down mechanism that reduces error weight sequentially. It avoids both the exponential overhead of lookup-table schemes and the requirement to perform faster correction with increasing system size, both present in the standard approach. The trickle-down QEC yields orders-of-magnitude lower logical error rates and reduction in qubit overhead, which we showed using dissipative repetition codes. 

More broadly, we can connect the trickle-down correction mechanism to approximate decoding strategies of topological codes as well as dissipative bosonic QEC mediated by auxiliary qubits~\cite{Note1}. In this sense, our construction, based on physically motivated principles akin to cooling, offers a unifying perspective applicable across the field of QEC and beyond such as cooling-like algorithms~\cite{raghunandan_initialization_2020,zapusek_scaling_2025}.

The main limitation of dissipative trickle-down QEC is its reliance on nonlocal operators, which is challenging for platforms with nearest-neighbor connectivity. However, this need not be restrictive: platforms with flexible or all-to-all connectivity (e.g., trapped-ion and neutral-atom arrays) are well suited to such operations, and codes with intrinsically nonlocal stabilizers, such as qLDPC codes, are under active investigation. Interesting directions for future works are the exploration of trickle-down schemes for measurement-feedback QEC and the development of generalized error-transparent~\cite{lebreuilly_autonomous_2021} or minimally distorting~\cite{rojkov_two-qubit_2024} Hamiltonians for logical operation of quantum codes~\cite{Note1}.

\medskip\noindent
\textit{Acknowledgements.---}
I.R. thanks Wojciech Adamczyk, Matteo Simoni, and Jonathan Home for insightful discussions on quantum error correction and its analogy to cooling, as well as for their valuable feedback on the manuscript. The authors acknowledge funding from the Swiss National Science Foundation (Ambizione Grant No.~PZ00P2$\_$186040) and the ETH Research Grant ETH-49~20-2.

\medskip\noindent
\textit{Note added.---} Following the release of our work, related schemes implementing some dissipative trickle-down correction have been proposed for surface codes~\cite{lake_local_2025,lake_fast_2025}.


\bibliography{references}

\newpage
\onecolumngrid
\begin{center}
    \large\bf End Matter
\end{center}
\twocolumngrid

\noindent
\textit{Nomenclature.} \\[5pt]
\begin{tabular}{ c l }
    $\msf{H},\,\msf{C},\,\msf{P}$ & \makecell{Hilbert \uline{spaces or subspaces}\\associated to codewords or\\to some projectors}\\[15pt]
    $\cl{L},\,\cl{R},\,\cl{K}_t$ & \uline{Superoperators} or \uline{CPTP maps}\\[5pt]
    $H,\,L,\,E,\,P,\,S,\,U$ & \makecell{\uline{Operators} representing \\Hamiltonians, jumps, etc.} \\[10pt]
    $\bb{L},\,\bb{E},\,\bb{P}$ & \makecell{\uline{Sets} of error operators,\\error subspaces or projectors}
\end{tabular}

\medskip\noindent
\textit{Probability of error and lookup-table correction jumps} ---
In the quantum trajectories picture~\cite{dalibard_wave-function_1992,molmer_monte_1993,carmichael_statistical_2008} of an incoherent dynamics, the occurrence of a quantum jump follows a Bernoulli process with a probability $\delta p = \delta t\,\sum_j \delta p_j$ where each $L_j$ occurs with a probability $\delta p_j / \delta p$ and ${\delta p_j= \delta t\,\langle L_j^\dagger L_j \rangle}$. For a system subject to $N_e$ unitary errors $L_{\n{e},i}$ occurring at a rate $\Gerr$ and a dissipative correction as in Eq.~\eqref{eq:jump_lookup_table}, the probability of any jump to occur at a time $t$ equals $\delta p =\delta t \left( N_\n{e} \Gerr + \Gqec\right)$. The probabilities that it corresponds to an error or a correction are
\be \label{eq:proba_err_corr_std}
    p_\n{e} = \frac{N_\n{e}\Gerr}{N_\n{e}\Gerr+\Gqec} 
    \quad\text{and}\quad 
    p_\n{c} = \frac{\Gqec}{N_\n{e}\Gerr+\Gqec}\,.
\ee
If the number of $L_{e,i}$ increases polynomially with the system size, the correction rate $\Gqec$ must scale as $\cl{O}(\n{poly}(n))$ to maintain a constant ratio $p_\n{e}/p_\n{c}$.
The limitations of the lookup-table dissipative QEC are discussed further in the Supplemental Material~\cite{Note1}.

\medskip\noindent
\textit{Number of projectors in each jump} ---
$N_\n{proj}$ follows a sum of binomial coefficients. This accounts for the fact that a logical state can transition to an erroneous state of order $m$ through any of the $\abs{\bb{L}}$ choose $m$ possible paths. The jumps in Eq.~\eqref{eq:jump_optimal} then coherently reduce the error weight by one, driving all $\abs{\bb{L}-1}$ choose $m$ paths that achieve this. Assuming ${\epsilon=\ell/\abs{\bb{L}}<1/2}$ and using Stirling's formula, $N_\n{proj}$ can be lower bounded by  
\be \label{eq:num_of_proj_bound}
\begin{split}
    N_\n{proj} 
    &\geq \frac{2^{\left(\abs{\bb{L}}-1\right)\,H(\epsilon)}}{\sqrt{8(\ell-1)\left(1-\frac{\ell-1}{\abs{\bb{L}}-1}\right)}} \\
    &\geq \frac{2^{4(\ell-1)\left(1-\frac{\ell-1}{\abs{\bb{L}}-1}\right)}}{\sqrt{8(\ell-1)\left(1-\frac{\ell-1}{\abs{\bb{L}}-1}\right)}} 
\end{split}
\ee
where $H(p)=-p\log_2(p)-(1-p)\log_2(1-p)\geq4p(1-p)$ is the binary entropy function.

\medskip\noindent
\textit{Effective dynamics of the trapped-ion model} ---
The presented trapped-ion scheme relies on resonance engineering~\cite{reiter_dissipative_2017} and the effective dynamics is found using the effective operator formalism~\cite{reiter_effective_2012}. First we introduce many-body basis involving computational states and excitations in modes $a$ and $b$
\be \label{eq:states_i1_ic}
\begin{split}
    &\ket{i_1,\ldots,i_c} = \\ &\quad \Bigg[\! 
    \left(\bigotimes_{k=1}^c \,\dyad{1}{0}_{i_k}\right)\! a^\dagger
    \!+ \left(\bigotimes_{k=1}^c \,\dyad{0}{1}_{i_k}\right)\! b^\dagger 
    \Bigg]\!\ketL{\psi}\!\ket{0,0}
\end{split}
\ee
where $\ketL{\psi}=\alpha\ketL{0} + \beta\ketL{1} = \alpha \ket{0}^{\otimes n} + \beta \ket{1}^{\otimes n}$, and $\ket{0,0}$ -- the ground state of both oscillators. $\ket{i_1,\ldots,i_c}$ represent the erroneous code states of weight $c \leq \ell$ with exactly one excitation in either of the two modes. The indices $\{i_1,\ldots,i_c\}$ denote the corrupted physical qubits. Moreover, an excitation in mode $a$ is associated to the error states decoded as logical zero $\ketL{0}$ while an excitation in mode $b$ -- to the error states decoded as $\ketL{1}$. Other important many-body states necessary for the understanding of the scheme are
\be \label{eq:states_Xi}
    \ket{\Xi_{i_1,\ldots,i_c}}=\frac{1}{\sqrt{c}}
    \left( \sum_{k=1}^{c} a \dyad{e}{1}_{i_k} \!+ b \dyad{f}{0}_{i_k} \!\right)\!
    \ket{i_1,\ldots,i_c}
\ee
involving ground states of both modes and ions' internal states $\ket{e}$ and $\ket{f}$. Assuming the bosonic mode to remain in the ground and first excited states only, the lowest order term of the Hamiltonian $H_0+H_2$ (where $H_0$ is the detuning part of $H_1$) in this new basis reads
\be \label{eq:hamilton_tilde}
\begin{split}
    \tilde{H} = \sum_{c=1}^{\ell} \sum_{i_1\neq\ldots\neq i_c}^{n} 
    &\Delta \dyad{\Xi_{i_1,\ldots,i_c}}{\Xi_{i_1,\ldots,i_c}} \\
    +&\delta \dyad{i_1,\ldots,i_c}{i_1,\ldots,i_c} \\
    +&\sqrt{c}\,G \dyad{i_1,\ldots,i_c}{\Xi_{i_1,\ldots,i_c}} + \n{H.c.} \,.
\end{split}
\ee
The \textit{tilde} indicates that it is only a portion of the complete Hamiltonian, it does not capture higher order processes involving two or more ions in the excited manifold. However, these processes will be out of resonance as happening at higher rates, e.g., the second order processes involving states with exactly two $\ket{e/f}$ will happen at rates $2\Delta$ and $\sqrt{c \choose 2} G$. The leading undesired process is the one that deteriorates an initial erroneous state by introducing one additional error. To describe it, we define the complementary many-body states $\ket{\overbar{i_1,\ldots,i_c}} = X_\n{L} \ket{i_1,\ldots,i_c}$ with $X_L$ --- the logical Pauli X operator. These states are associating an excitation in $a$ to error states decoded as logical one $\ketL{1}$ and an excitation in $b$ -- to error states decoded as $\ketL{0}$. The last many-body states that we require are
\be 
\begin{split}
    &\ket{\chi_{i_1,\ldots,i_c}}=\frac{1}{\sqrt{n-c}} \\
    &\quad\left( \sum_{i_k \notin \{i_1,\ldots,i_c\} } a \dyad{e}{1}_{i_k} + b \dyad{f}{0}_{i_k} \right) 
    \ket{\overbar{i_1,\ldots,i_c}}
\end{split}
\ee
where the sum is taken over all non erroneous qubits of the state $\ket{i_1,\ldots,i_c}$. These states are the undesired counterparts of $\ket{\Xi_{i_1,\ldots,i_c}}$ from Eq.~\eqref{eq:states_Xi}. The lowest order undesired Hamiltonian associated to $H_0+H_2$ then reads
\be \label{eq:hamilton_u_tilde}
\begin{split}
    \tilde{H}_u = \sum_{c=1}^{\ell} \sum_{i_1\neq\ldots\neq i_c}^{n}
    &\,\Delta \dyad{\chi_{i_1,\ldots,i_c}}{\chi_{i_1,\ldots,i_c}} \\
    +&\,\delta \dyad{\overbar{i_1,\ldots,i_c}}{\overbar{i_1,\ldots,i_c}} \\ 
    +&\,\sqrt{n-c}\,G \dyad{\overbar{i_1,\ldots,i_c}}{\chi_{i_1,\ldots,i_c}} \,.
\end{split}
\ee

The Hamiltonian $H_3$ represents single addressing lasers that couple each ion at position $j$ to an auxiliary motional mode $c_j$ which in turn is sympathetically cooled. In the regime $\kappa \gg g$, the auxiliary mode can be adiabatically eliminated and approximate the whole process as jump operator acting on every ion individually 
\be \label{eq:sympa_cooling}
    L_{\n{eng},j}=\sqrt{\kappa_\n{eng}}\big( \dyad{0}{e}_j + \dyad{1}{f}_j \big)
\ee
with $\kappa_\n{eng} \propto g^{2}/\kappa$.

We derive the effective Liouvillian that governs the many-body dynamics using the effective operator formalism~\cite{reiter_effective_2012}. Assuming that the ground subspace of the total Hilbert space is spanned by $\{\ket{0},\ket{1}\}^{\otimes n}\otimes\ket{0,0}$, while the excited manifold is spanned by states containing at most one excited state, which could be $\ket{e}$, $\ket{f}$, $\ket{1,0}$ or $\ket{0,1}$, the non-Hermitian Hamiltonian reads
\be
\begin{split}
    H_\n{NH} &= H_{0} + H_{2} + H_{3} - \frac{i}{2} \sum_{j=1}^n L_{j}^{\dagger} L_{j} \\
    &\approx H_{0} + H_{2} - \frac{i}{2} \sum_{j=1}^n L_{\n{eng},j}^{\dagger} L_{\n{eng},j} \\
    &\approx \tilde{H} + \tilde{H}_{u} - \frac{i}{2} \sum_{j=1}^n L_{\n{eng},j}^{\dagger} L_{\n{eng},j} =: \tilde{H}_\n{NH}
\end{split}
\ee
where in the first approximation we use Eq.~\eqref{eq:sympa_cooling}, and in the second Eqs.~\eqref{eq:hamilton_tilde} and~\eqref{eq:hamilton_u_tilde}. The last term can be rewritten in terms of  $\ket{\Xi_{i, \ldots i_{c}}}$ and $\ket{\chi_{i_1,\ldots,i_c}}$ in a similar manner as $H_0$ resulting in an energy shift of $-i\kappa_\n{eng}/2$. Thanks to the block form of $\tilde{H}_\n{NH}$, we invert every block individually
\be \label{eq:inv_NH_hamilton_reduced}
\begin{split}
    \left[\tilde{H}_\n{NH}\right]^{-1}\!\!=\sum_{c=1}^{\ell} \sum_{i_{i} \neq \ldots \neq i_{c}}^{n} 
    \frac{1}{\Delta'(c)}  &\dyad{\Xi_{i, \ldots i_{c}}}{\Xi_{i_{1} \ldots i_{c}}} \\
    + \frac{1}{\Delta'(n-c)} &\dyad{\chi_{i_1,\ldots,i_c}}{\chi_{i_1,\ldots,i_c}} + \ldots
\end{split}
\ee
For conciseness, we present here only the relevant part of~$\big[\tilde{H}_\n{NH}\big]^{-1}$. The rest of the terms are found in Supplementary Materials~\cite{Note1}. The scalars in Eq.~\eqref{eq:inv_NH_hamilton_reduced} are defined as
\be
    \Delta'(x) = \Delta - \frac{i}{2} \kappa_\n{eng} - \frac{x\,G^2}{\delta}
\ee
with $x\in\{1,\ldots,\ell\}$. The behavior of these functions in terms of the integer $x$ is the core of the QEC process.
 
To derive the effective jump operators we express the drive part $V := V_+ + V_{-}$ of the Hamiltonian $H_1$ in terms of our many-body states, which to the lowest order reads 
\be
\begin{split}
    \tilde{V}_+ = \tilde{V}_{-}^\dagger = 
    \sum_{c=1}^\ell \sum_{i_{i} \neq \ldots \neq i_{c}}^{n}
    \sqrt{c}\,&\frac{\Omega}{2} \dyad{\Xi_{i_1,\ldots,i_c}}{i_1,\ldots,i_c}_0 \\
    + \sqrt{n-c}\,&\frac{\Omega}{2} \dyad{\chi_{i_1,\ldots,i_c}}{i_1,\ldots,i_c}_0
\end{split}
\ee
where $\ket{i_1,\ldots,i_c}_0$ is identical to $\ket{i_1,\ldots,i_c}$ but with zero excitation in modes $a$ and $b$. These states correspond effectively to error states of weight $c$ of the repetition code. Assuming this drive to be weak such that we can treat it as a perturbative excitation of the ground subspace, the effective jump operators become
\[
\begin{split}
    &L_{\n{eff},j} = L_{\n{eng},j} \left[\tilde{H}_\n{NH}\right]^{-1} \tilde{V}_{+}\\
    = &\sum_{c=1}^\ell \sum_{i_{i} \neq \ldots \neq i_{c-1} \neq j}^{n} \sqrt{\kappa_\n{eff}(c)} 
    \dyad{i_1,\ldots,i_{c-1}}{i_1,\ldots,i_c}_0 \\
    +&\sum_{i_{i} \neq \ldots \neq i_{c} \neq j}^{n} \sqrt{\kappa_\n{eff}(n-c)}
    \dyad{i_1,\ldots,i_c,i_{c+1}=j}{i_1,\ldots,i_c}_0
\end{split}
\]
These jump operators consist of two processes: a desired one that reduces the weight of error states from $c$ to $c-1$, and an undesired one that increases the error weight by~1. The desired process, represented by the first sum, corrects all error states in which the $j$-th ion is erroneous. In contrast, the undesired process flips the state of the $j$-th ion even when it is not erroneous, thereby introducing an additional error. Rewriting $L_{\n{eff},j}$ in terms of projectors into the error subspaces yields the expression in Eq.~\eqref{eq:eff_jump_op_proj}. The effective rate $\kappa_\n{eff}(x)$ becomes
\be \label{eq:eff_jump_op_rate_deriv}
    \kappa_\n{eff}(x) = \kappa_\n{eng} \,\frac{\Omega^2}{4} \big\lvert\Delta'(x)\big\rvert^{-2}
\ee
and equals exactly the expression in Eq.~\eqref{eq:eff_jump_op_rate}. Fixing the coupling in the Hamiltonian $H_2$ to $G=G_0=\sqrt{\Delta\delta/x_0}$ with an arbitrary $x_0\in\{1,\ldots,\ell\}$ reduces Eq.~\eqref{eq:eff_jump_op_rate_deriv} to
\be \label{eq:eff_jump_op_rate_res_cond}
    \kappa_\n{eff}(x,x_0) = \frac{\Omega^2}{\kappa_\n{eng}} 
    \left[ \left(\frac{2\Delta}{\kappa_\n{eng} x_0}\right)^2 \big(x-x_0\big)^2 + 1 \right]^{-1}\!.
\ee
Both Eqs.~\eqref{eq:eff_jump_op_rate} and~\eqref{eq:eff_jump_op_rate_res_cond} describe a Lorentzian function of the integer parameter $x$ centered at $x_0$. Therefore, choosing the resonance condition $G=G_0$ favors the desired process $x_0\rightarrow x_0-1$, i.e. the correction of error states of weight $x_0$ towards errors states of weight $x_0-1$. The scale and height of the Lorentzian functions are defined as $x_0 \kappa_\n{eng}/2\Delta$ and $\Omega^2/\kappa_\n{eng}$, respectively. See further details in the Supplemental Material~\cite{Note1}.

\iftoggle{arXiv}{
\cleardoublepage 
\title{Supplementary Materials:\\Scalable dissipative quantum error correction for qubit codes}
\maketitle

\onecolumngrid
\setcounter{equation}{0}
\renewcommand{\theequation}{S\arabic{equation}}
\setcounter{figure}{0}
\renewcommand{\thefigure}{S\arabic{figure}}
\setcounter{page}{1}

\vspace*{-20pt}

\SMsec{Note on Knill--Laflamme condition}{sec:intro}

The general form of the Knill--Laflamme condition for a code $\msf{C}$ and a set of error operators $\bb{E}=\{E_i\}$ reads~\cite{knill_theory_1997}
\be
  P_\msf{C}E_j^\dagger E_i P_\msf{C}=\alpha_{ij}P_\msf{C} 
  \qquad \forall \,\, i,j
\ee
with $\alpha_{ij}$ being a scalar and $P_\msf{C}$ the projector on the logical subspace. This condition ensures that there exist a partitioning of the Hilbert space $\msf{H}$ into subspaces, each associated with an individual invertible (or approximately invertible~\cite{leung_approximate_1997}) error process $\tilde{E}_i$. These new error operators are obtained by unitarily transforming the set $\{E_i\}$ such that $\alpha_{ij}=p_i\delta_{ij}$ with $p_i$ quantifying the error detection probability of $\tilde{E}_i$. We can therefore consider without loss of generality that $\tilde{E}_i\equiv{E}_i$ what we do in the main text.

\SMsec{Lookup-table dissipative QEC}{sec:lookup_table}

In this section, we discuss in more details the two fundamental limitations of the standard dissipative QEC approach. Recall that the general form of the lookup-table Liouvillian is $\cl{L}_\n{c}(\rho)=\Gqec(\cl{R}(\rho) - \rho)$ with $\cl{R}$ being the recovery map. Correction jump operators satisfying this Liouvillian are given for example by Eq.~(1). These are not unique given the unitary invariance property of the Lindblad equation, however Eq.~(1) represents a convenient construction since for a large class of quantum codes, projectors $P_i$ onto error subspaces can be efficiently determined. For $\llbracket n,k,d \rrbracket$ stabilizer codes, for example, these are can be constructed using the generators $\big\{S_j\big\}_{i=1}^{n-k}$ of the stabilizer group, 
\be \label{eq:stabilizer_proj}
    P_i = P(\vec{s}_i) := \,\prod_{j=1}^{n-k}\, \frac{1}{2}\Big(1+ (-1)^{\vec{s}_i[j]} \, S_j \Big)\,,
\ee
where ${\vec{s}_i\in\{0,1\}^{n-k}}$ and $\vec{s}_i[j]$ the $j$-th element. The set $\{L_{\n{c},i}\}$ results in $N_\n{sub}=2^{n-k}-1$ jump operators. For non-stabilizer codes, a similar approach can be employed by introducing additional preventive jump operators that map each state in the orthonormal basis spanning the uncorrectable subspace into the correctable subspace~\cite{lihm_implementation_2018,lebreuilly_autonomous_2021}. However, in this case, the number of required jump operators also scales exponentially with the system size.

One of the most studied dissipative QEC codes is the three-qubit bit-flip code~\cite{paz_continuous_1998,ahn_continuous_2002,sarovar_continuous_2005,ippoliti_perturbative_2015,reiter_dissipative_2017,lihm_implementation_2018}. In this setting, a single logical qubit is encoded in a system of three qubits using ${\msf{C}=\n{span}\{\ket{000},\ket{111}\}}$. It is protected against single qubit bit flips $\{{L_{\n{e},i}=\sqrt{\Gerr}\,X_i}\}_{i=1}^{3}$, i.e. single-qubit Pauli $X$ errors at a rate $\Gerr$, using
\be \label{eq:jump_rep3}
	L_{\n{c},i} = \sqrt{\Gqec}\,X_i\,
    \frac{1-Z_iZ_j}{2}\,\frac{1-Z_iZ_l}{2}
\ee
where $i,j,l \in \{1,2,3\}$. This operator effectively flips the $i$th qubit at a rate $\Gqec$ if and only if its parity disagrees with the one of the two other qubits. In this example, $X_i$ corresponds to the unitary $U_i^\dagger$ from Eq.~\eqref{eq:jump_lookup_table} while the projector $P_i$ is built using operators $Z_iZ_j$ and $Z_iZ_l$ querying the parity of pairs of qubits. This construction and methods for its implementation has been generalized to other stabilizer codes in Refs.~\cite{hsu_method_2016,oreshkov_chapter8_2013} using among others Eq.~\eqref{eq:stabilizer_proj}.

\SMsec{Limitations of the lookup-table dissipative QEC}{sec:limits_lookup_table}

The lookup-table dissipative QEC has two fundamental limitations. First, the number of required correction operators scales exponentially with the number of qubits $n$. Even though each $L_{\n{c},i}$ can be implemented using $\cl{O}(n)$ auxiliary qubits allowing a tradeoff between space and time complexity, the exponential operational cost is unavoidable~\cite{heusen_measurement-free_2024}. The same limitation appears in measurement-feedback QEC, where the construction in Eq.~\eqref{eq:jump_lookup_table} reduces to a lookup-table decoder whose size grows exponentially with code size. Approximate decoders avoid this by incorporating error models of the system, e.g. the locality of errors, the qubit connectivity and the device calibrations, reducing effectively the number of error subspaces to be corrected. Second, as mentioned in the main text and the End Matter, even if the number of errors increases only polynomially with system size, the correction rate $\Gqec$ must scale accordingly (i.e. $\cl{O}(\n{poly}(n))$) to maintain a constant ratio $p_\n{e}/p_\n{c}$ between the probabilities of error and correction jumps. This has been previously noted in Refs.~\cite{pastawski_quantum_2011,shtanko_bounds_2025}. In the latter work, authors conlcuded that for faster-than-polynomial reduction of the logical error rate, $\Gqec$ must scale as $\cl{O}(\log(n))$. In other word, solely incresing the code size does not help to reduce the logical error rate, one must also perform the error-check and correction operations equivalently faster.

These limitations are intrinsic to lookup-table Liouvillians $\cl{L}_\n{c}(\rho)=\Gqec(\cl{R}(\rho) - \rho)$, which restore the codespace only asymptotically, while having little effect at intermediate times. To see this, note that $\cl{L}_\n{c}$ generates a time evolution described by
\be
  \rho(t) = e^{-\Gqec t} \rho(0) + (1-e^{-\Gqec t}) \cl{R}(\rho(0))
\ee
which ensures that the system converges to the codespace. Consequently, the trace distance between $\rho(t)$ and the steady state $\cl{R}(\rho(0))$ evolves independently of the code $\msf{C}$,
\be
  \norm{\rho(t)-\cl{R}(\rho(0))}_1=e^{-\Gqec t}\norm{\rho(0)-\cl{R}(\rho(0))}_1.
\ee
In order to have a faster convergence in the distance, a correction Liouvillian should instead generate a time evolution of the form~\cite{shtanko_bounds_2025}
\be \label{eq:liouv_gen_decoder2}
    \rho(t) = e^{-\Gqec t}\,\cl{K}_t(\rho(0)) + (1-e^{-\Gqec t}) \cl{R}(\rho(0))
\ee
with $\cl{K}_t$ being a time-dependent CPTP map with a nontrivial action on the error subspaces. When 
\be 
  \norm{\cl{K}_{t}(\rho(0))-\cl{R}(\rho(0))}_1<\norm{\cl{K}_{t'}(\rho(0))-\cl{R}(\rho(0))}_1
\ee
for ${t<t'}$, we can think of it as a partial recovery map~\cite{shtanko_bounds_2025} which reduces the error weight without fully recovering the codespace. \citet{pastawski_quantum_2011} demonstrated that such Liouvillians exist by proposing a construction where $\cl{L}_\n{c}$ is a sum of recovery operations acting on different stages of a concatenated QEC code whose strength decay as a power law with the level concatenation, i.e. 
\be \label{eq:liouv_pastawski}
    \cl{L}_\n{c}(\rho)= \sum_i \Gqec^{(i)} (\cl{R}_i(\rho) - \rho)
\ee 
where $\Gqec^{(i)} \sim \cl{O}\big(\Gamma^i\big)$ is the correction rate at the concatenation level $i$ ($i=1$ being the physical level) and $\Gamma<1$ -- a constant. Although Ref.~\cite{pastawski_quantum_2011} showed that partial recovery maps for dissipative QEC can be constructed when one considers specific concatenation of $\llbracket n,3,1 \rrbracket$ codes, it remains unclear how to design such Liouvillians for general codes, how they perform against the lookup-table solution, and, crucially, how many corrective jump operators must be engineered and what are the necessary resources. These are the main questions we address in this work.

\SMsec{Trickle-down QEC of repetition codes}{sec:repetition_code}

In this section, we provide additional details about the trickle-down dissipative QEC of repetition codes studied in the main text. Let us first comment our trickle-down construction in general. We illutstrate it in Fig.~\ref{fig:tricle_down_diagram}. Error subspaces are grouped by their error weight $j$, with each horizontal block $\bb{P}_j$ representing all subspaces at weight $j$. The number of subspaces per level increases with the number of fundamental noise processes. Arrows indicate the action of trickle-down jump operators, where identical arrow styles denote transitions implemented by the same operator. The Knill--Laflamme condition in Eq.~(2) determines which subspace pairs can be corrected by a common operator. The correction jump operators are then constructed using Eq.~(4). For the repetition code, the projectors onto the error subspaces follow Eq.~\eqref{eq:stabilizer_proj} and use the generators of the code's stabilizer group, $\{Z_1 Z_2,Z_2 Z_3, ..., Z_{n-1} Z_n\}$. Alternatively, we can also define these projectors using $P_\msf{C}$ and the error operators. For example, for single qubit bit-flip subspaces the $n$ projectors are given by $P_i=X_i P_\msf{C} X_i \,\in \bb{P}_1$. The projectors onto higher order error subspaces are obtained in a similar way. This way the desired jump operators from Eq.~(4) can also be written as
\be \label{eq:jump_optimal_rep}
    L_{\n{c},i}^{(1)} = \sqrt{\Gqec} \sum_{j=1}^{\ell} \sum_{k=1}^{K_j} P_{i,k}^{(j-1)} X_i \quad \forall i \in \{1,\ldots,n\}
\ee 
using the fact that the recovery unitaries $U_i^{(1)\dagger}$ are exactly the bit-flip operators $X_i$ thanks to their involutory character, $X_i^2=\id$. 

To study the correction mechanism of the jump operators in Eqs.~(1) and~(4), we first simulate the dynamics of a 11-qubit repetition code initialized in the state $\ket{\psi_\n{L}}\propto\ket{0_\n{L}}+i\ket{1_\n{L}}$ and subjected to bit-flip error ${L_{\n{e},i}=\sqrt{\Gerr}\,X_i}$ and correction jumps. After an evolution time $t<3/\Gerr$ we decode the system state by majority voting and compute the logical fidelity with the initial state. The results are presented in Fig.~\ref{fig:td_vs_lt_infidelity}. In addition, to simulate the optimal correction operators, we also perform simulations with jump operators correcting errors only up to orders ${m<\ell=5}$ (i.e. we build the jump operators in Eq.~\eqref{eq:jump_optimal_rep} with the first sum going only to $m$). At short timescales all infidelity curves follow an identical trend dictated by the final decoding step of the simulation using majority vote. This decay of the fidelity described by $\cl{O}\!\left(t^{\,\ell+1}\right)$ comes from the unavoidable probability that multiple errors arise and accumulate resulting in an error of weight at least $\ell+1$ and a misdecoding of the system state. Thus, majority vote alone allows a short timescales to slow down the loss of information about the initial logical state by using codes with larger $\ell$. The goal of an error correction protocol is however to (1) reduce the logical error rate at longer timescales, and (2) start this secondary regime at the earliest time possible. A clear example of this is the infidelity evolution using the lookup-table correction jump operators (in black in Fig.~\ref{fig:td_vs_lt_infidelity}). We observe that for $t\gtrsim0.05/\Gerr$ the logical infidelity starts to increase at a much slower rate than for earlier times such that at $3/\Gerr$ the infidelity is only $\sim10^{-4}$ as opposed to $0.5$ in the uncorrected case. We now compare this to the performance of our trickle-down solution. When considering correction jump operators with projectors up to error weight ${m<4}$, the logical fidelity is lower than for the lookup-table solution. This is expected as the error processes are not fully corrected resulting in always growing uncorrectable error subspace. However, when $m=4$ and $m=\ell$, our trickle-down solution outperforms the lookup-table one and, in the optimal case, by more than an order of magnitude. We also notice that for $m=\ell$, the onset of the slow increase of the logical infidelity is shifted to earlier timescales by a factor of $5$ which is a signature that correcting errors in a trickle down fashion enhances the probability that a correction jump occurs. In other words, it enhances the ratio $p_\n{e}/p_\n{c}$.

\begin{figure*}[t!]
    {\phantomsubcaption\label{fig:tricle_down_diagram}}
    {\phantomsubcaption\label{fig:td_vs_lt_infidelity}}
    \includegraphics{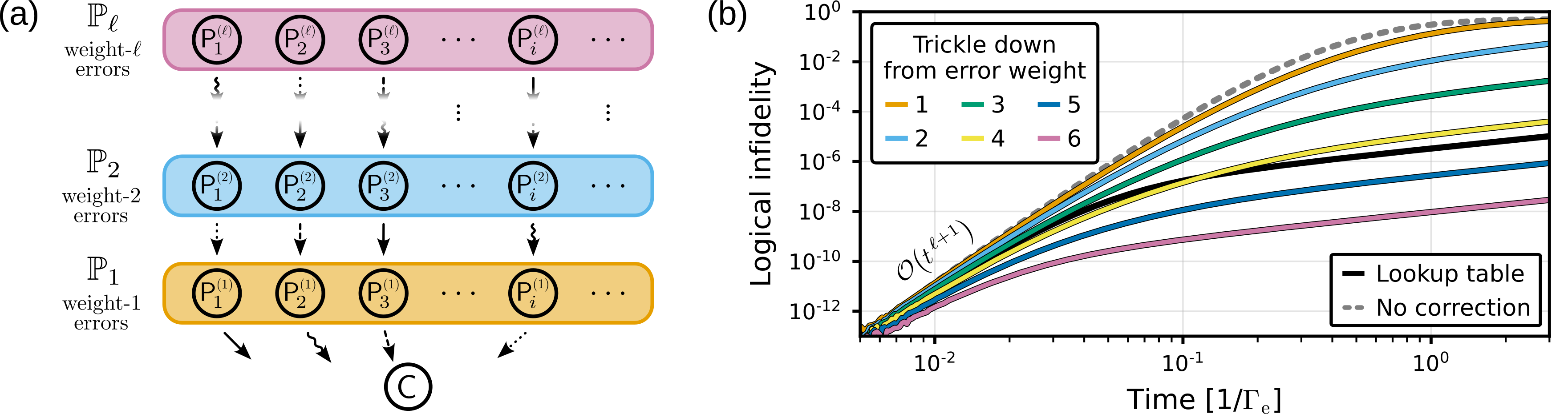}
    \caption{\textit{Trickle-down QEC.} (a) Schematic of the trickle-down dissipative error correction scheme. Error subspaces $\msf{P}_i^{(j)}$ are grouped by their error weight $j$, with each horizontal block $\bb{P}_j$ representing all subspaces at weight $j$. The number of subspaces per level increases with the number of fundamental noise processes. Error weights range up to $\ell$, the maximum correctable weight. Arrows indicate the action of trickle-down jump operators, where identical arrow styles denote transitions implemented by the same operator. The Knill--Laflamme condition in Eq.~\eqref{eq:KL_error_subspaces} determines which subspace pairs can be corrected by a common operator. (b) Logical infidelity over time for a 13-qubit repetition code initialized in $\ketL{\psi} \propto \ketL{0} + i\ketL{1}$, subject to bit-flip noise (with error rate $\Gerr=10^{-2}\Gqec$) and dissipative QEC. We compare the performance of the lookup-table approach and the trickle-down scheme, the latter is performed up to different error-weight levels.}
    \label{fig2}
\end{figure*}

As mentioned in the main text, the long timescale trend of the infidelity curves can be described by the function $F(t)=0.5\left[1+\exp(\epsilon\,(t-\tau))\right]$ where $\epsilon$ represents the rate of the decay and $\tau$ the time at which one switches from one regime to another. As explained, we extract and process these parameters by fitting this function to time evolutions shown in Fig.~\ref{fig:td_vs_lt_infidelity}. In other words, each data point in Fig.~2a is obtained from a separate time evolution simulation.

Finally, we note that our simulations in Fig.~2 for the lookup table solution aligns with the most recent experimental value of $\Lambda$ obtained for a measurement-feedback repetition code implemented on superconducting-qubit device~\cite{acharya_quantum_2024}. In this work, the authors obtained an exponential suppression factor of $8.4\,\pm\,0.1$ for an averaged physical error probability slightly below $10^{-2}$. These values are comparable with our results from Fig.~2.

\begin{figure*}[b!]
    {\phantomsubcaption\label{fig:trapped_ion_scheme}}
    {\phantomsubcaption\label{fig:effective_rates}}
    \includegraphics{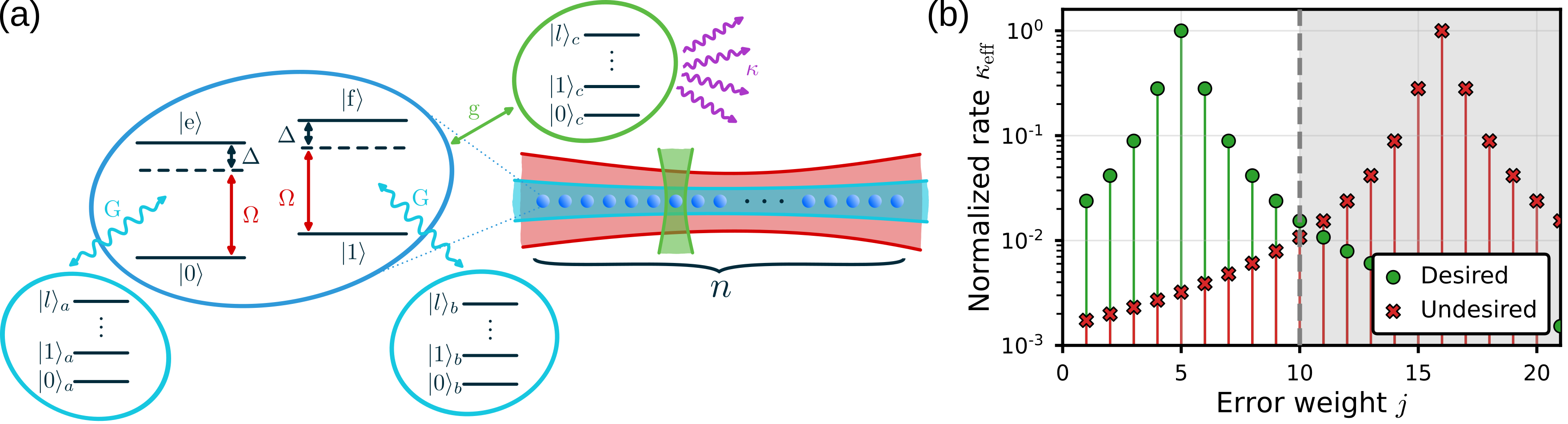}
    \caption{\textit{Trapped-ion implementation.} (a) Schematic of a trapped-ion setup for implementing trickle-down correction of a repetition code. Each of the $n$ qubits is encoded in two electronic levels of a single ion. The correction mechanism uses two additional electronic levels $\ket{e}$ and $\ket{f}$, two global motional modes $a$ and $b$, and a local mode $c$ that is sympathetically cooled to remove entropy. The process is driven by two global laser beams implementing the interactions in Eqs.~\eqref{eq:hamilton1} and~\eqref{eq:hamilton2}, and local beams implementing Eq.~\eqref{eq:hamilton3}. (b) Effective rates of the jump operators from Eq.~(6) for a $21$-qubit repetition code, shown as a function of the error weight of the target subspaces. Desired processes reduce the error weight via $j \rightarrow j-1$ transitions, while undesired ones increase it via $j \rightarrow j+1$. Shaded regions correspond to processes beyond the correctable error weight $\ell = 10$, which are inactive but included to highlight the symmetry of the Lorentzian profiles.} 
    \label{fig4}
\end{figure*}

\SMsec{Trapped-ion implementation of dissipative repetition code}{sec:trapped_ion_rep_code}

In this section, we provide additional details about the trapped-ion implementation of the dissipative repetition code. Our scheme is schematically illustrated in Fig.~\ref{fig:trapped_ion_scheme}. As explained in the main text, $H$ consists of three parts which we restate here. First, a global laser drive addressing simultaneously $|0\rangle \leftrightarrow|e\rangle$ and $|1\rangle \leftrightarrow|f\rangle$ with a strength $\Omega$ and detuned by $\Delta$ from the resonance
\be \label{eq:hamilton1}
    H_1 = \sum_{i=1}^{n}  \Delta \left( \dyad{e}{e}_i + \dyad{f}{f}_i \right) + 
    \frac{\Omega}{2} \left( \dyad{e}{0}_i + \dyad{f}{1}_i \right) + \n{H.c.}
\ee 
which we can also split into the sum of a detuning $H_0$ and perturbation $V$ parts. Next, we require a second global laser beam now coupling $\ket{0}/\ket{f}$ and $\ket{1}/\ket{e}$ to the motional modes $a$ and $b$, respectively, with a strength $G$ and a detuning $\delta$,
\be \label{eq:hamilton2}
    H_2 = \delta \left( a^\dagger a + b^\dagger b \right) + G \sum_{i=1}^{n} a^\dagger \dyad{1}{e}_i + b^\dagger \dyad{0}{f}_i + \n{H.c.}
\ee
Third, we add a single addressing laser coupling an ion at position $i$ to an auxiliary motional mode $c_i$ 
\be \label{eq:hamilton3}
    H_{3,i} = g\,c_i^{\dagger} \left( \dyad{0}{e}_i + \dyad{1}{f}_i \right) + \n{H.c.}
\ee
We use this mode to remove entropy from the system through jump operators $L_{i}=\sqrt{\kappa} \, c_i$ with a tunable rate $\kappa$. The error correction mechanism behind the total Hamiltonian $H=H_1+H_2+H_3$ is the following: using $H_2$ in a strong-driving regime creates many-body resonances between computational states and superpositions of states with a single excitation $\ket{e}$ or $\ket{f}$. Driving $H_1$ in a perturbative fashion and carefully setting both detunings $\Delta$ and $\delta$, we can address specific transitions between those states.

\medskip
Let us now present further explanations regarding our derivation from the End Matter. We start directly with the inversion of the non-Hermitian Hamiltonian $\tilde{H}_\n{NH}$ in Eq.~(16) which in the lowest order reads
\be \label{eq:inv_NH_hamilton}
\begin{split}
    \left[\tilde{H}_\n{NH}\right]^{-1}\!=\sum_{c=1}^{\ell} \sum_{i_{i} \neq \ldots \neq i_{c}}^{n} \,\,\, &
    \frac{1}{\Delta'(c)}  \dyad{\Xi_{i, \ldots i_{c}}}{\Xi_{i_{1} \ldots i_{c}}} +
    \frac{1}{\Delta'(n-c)} \dyad{\chi_{i_1,\ldots,i_c}}{\chi_{i_1,\ldots,i_c}} + \\
    + & \frac{1}{\delta'(c)}  \dyad{i_1,\ldots,i_c}{i_1,\ldots,i_c} + 
    \frac{1}{\delta'(n-c)}\dyad{\overbar{i_1,\ldots,i_c}}{\overbar{i_1,\ldots,i_c}} + \\
    + & \frac{1}{G'(c)} \dyad{i_1,\ldots,i_c}{\Xi_{i_1,\ldots,i_c}} +
    \frac{1}{G'(n-c)}  \dyad{\overbar{i_1,\ldots,i_c}}{\chi_{i_1,\ldots,i_c}} +\n{H.c.}
\end{split}
\ee
The scalar parameters in Eq.~\eqref{eq:inv_NH_hamilton} are defined with the functions
\be
\begin{split}
    \Delta'(x) = \Delta - \frac{i}{2} \kappa_\n{eng} - \frac{x\,G^2}{\delta} \qquad\quad
    \delta'(x) = \delta - \frac{x\,G^2}{\Delta - \frac{i}{2}\kappa_\n{eng}} \qquad\quad
    G'(x) = \sqrt{x}\,G \left(1 - \delta\, \frac{\Delta - \frac{i}{2}\kappa_\n{eng}}{x\,G^2}  \right) 
\end{split}
\ee
where $x\in\{1,\ldots,\ell\}$.

To derive the effective jump operators we must express the drive part $V := V_+ + V_{-}$ of the Hamiltonian $H_1$ in Eq.~\eqref{eq:hamilton1} in terms of our many-body states. In the lowest order, the drive reads 
\be
    V = V_+ + V_{-} \qquad\n{with}\qquad \tilde{V}_+ = \tilde{V}_{-}^\dagger = \sum_{c=1}^\ell \sum_{i_{i} \neq \ldots \neq i_{c}}^{n}
    \sqrt{c}\,\frac{\Omega}{2} \dyad{\Xi_{i_1,\ldots,i_c}}{i_1,\ldots,i_c}_0 + \sqrt{n-c}\,\frac{\Omega}{2} \dyad{\chi_{i_1,\ldots,i_c}}{i_1,\ldots,i_c}_0
\ee
where we define $\ket{i_1,\ldots,i_c}_0$ as the same state $\ket{i_1,\ldots,i_c}$ but with zero excitation in oscillators $a$ and $b$. These states are effectively representing error states of weight $c$ of the repetition code. We consider this drive to be weak such that one can treat it as a perturbative excitation of the ground subspace of the system. Following Ref.~\cite{reiter_effective_2012}, the effective jump operators become
\be \label{eq:eff_jump_op}
\begin{split}
    L_{\n{eff},j} = L_\n{eng,j} \left[\tilde{H}_\n{NH}\right]^{-1} \tilde{V}_{+}
    &= \sum_{c=1}^\ell \,\,\sum_{i_{i} \neq \ldots \neq i_{c-1} \neq j}^{n} 
    \sqrt{\kappa_\n{eng}}\,\frac{\Omega}{2} \frac{1}{\Delta'(c)} \,
    \underbrace{\ket{i_1,\ldots,i_{c-1}}_0}_{(c-1)\,\n{errors}}\!\underbrace{\bra{i_1,\ldots,i_c}_0}_{(c)\,\n{errors}} + \\
    &\,+ \sum_{c=1}^\ell \sum_{i_{i} \neq \ldots \neq i_{c} \neq j}^{n} 
    \sqrt{\kappa_\n{eng}}\,\frac{\Omega}{2} \frac{1}{\Delta'(n-c)} \,
    \underbrace{\ket{i_1,\ldots,i_c,i_{c+1}=j}_0}_{(c+1)\,\n{errors}}\!\underbrace{\bra{i_1,\ldots,i_c}_0}_{(c)\,\n{errors}}\,.
\end{split}
\ee
These jump operators consist of two processes: a desired one that reduces the weight of error states from $c$ to $c-1$, and an undesired one that increases the error weight by 1. The desired process, represented by the first sum, corrects all error states in which the $j$-th ion is erroneous. In contrast, the undesired process flips the state of the $j$-th ion even when it is not erroneous, thereby introducing an additional error. Rewriting the jump operator in terms of projectors into the error subspaces of weight $c$ gives
\be \label{eq:eff_jump_op_proj_2}
    L_{\n{eff},j}
    =\sum_{c=1}^\ell \sqrt{\kappa_\n{eff}(c)} 
    \left(\dyad{0}{1}_j P_{1,c}+\dyad{1}{0}_j P_{0,c}\right)+ 
    \sqrt{\kappa_\n{eff}(n-c)} 
    \left(\dyad{1}{0}_j P_{1,c}+\dyad{0}{1}_j P_{0,c}\right)
\ee
where the projectors $P_{0,c}$ and $P_{1,c}$ are defined as 
\be
    P_{\psi,c} := \sum_{i_{i} \neq \ldots \neq i_{c}}^{n} 
    \left(\bigotimes_{k=1}^c \,X_{i_k}\right) \dyad{\psi}{\psi}_L \left(\bigotimes_{k=1}^c \,X_{i_k}\right)\,.
\ee
The effective rate function $\kappa_\n{eff}(x)$ is obtained using
\be \label{eq:eff_rate}
    \kappa_\n{eff}(x) = 
    \kappa_\n{eng} \,\frac{\Omega^2}{4} \big\lvert\Delta'(x)\big\rvert^{-2} = 
    \kappa_\n{eng} \, \frac{\Omega^2}{4} \left[ \left(\Delta - \frac{x G}{\delta}\right)^2 + \frac{1}{4} \kappa_\n{eng}^2 \right]^{-1}
\ee
Fixing the coupling in the Hamiltonian $H_2$ to $G=G_0=\sqrt{\Delta\delta/x_0}$ with an arbitrary $x_0\in\{1,\ldots,\ell\}$ reduces the previous expression to
\be \label{eq:eff_rate_res_cond}
    \kappa_\n{eff}(x,x_0) = 
    \frac{\Omega^2}{\kappa_\n{eng}} \left[ \left(\frac{2\Delta}{\kappa_\n{eng} x_0}\right)^2 \big(x-x_0\big)^2 + 1 \right]^{-1}
    \,.
\ee
Both Eqs.~\eqref{eq:eff_rate} and~\eqref{eq:eff_rate_res_cond} describe a Lorentzian function of the integer parameter $x$ centered at $x_0$. Therefore, choosing the resonance condition $G=G_0$ favors the desired process $x_0\rightarrow x_0-1$, i.e. the correction of error states of weight $x_0$ towards errors states of weight $x_0-1$. The scale and height of the Lorentzian functions are defined as $x_0 \kappa_\n{eng}/2\Delta$ and $\Omega^2/\kappa_\n{eng}$, respectively. The normalized rates are represented in Fig.~\ref{fig:effective_rates} for a system of $n=21$ qubits, $x_0=5$, $\Delta=4\kappa_\n{eng}$, $\delta=3.46\kappa_\n{eng}$ and $G=1.66\kappa_\n{eng}$. The shaded area represents system states that have more errors than the code distance and are taken into account in undesired process. Although irrelevant for the later discussion, it is interesting to note that as expected the effective rates of the undesired process (red bars in the unshaded area) simply corresponds to the reflection of the Lorentzian function of the desired process past the code distance $d$ (green bars in the unshaded area).

The whole analysis above did not take into account one last undesired term in the the first order expansion $\tilde{H}_u$ given in Eq.~\eqref{eq:hamilton_u_tilde}, which is the process that adds a single error to the logical states themselves. Fortunately, this process can be easily included if we allow the abuse of notation $\ket{i_1,\ldots,i_c}$ with $c=0$ to represent the logical states $\ketL{0}$ and $\ketL{1}$. In this case the effective jump operator representing the $j$-th ion will be identical to the expressions in Eqs.~\eqref{eq:eff_jump_op} and \eqref{eq:eff_jump_op_proj_2} but with the sum of the undesired process starting at $c=0$.

The most important feature of the proposed dissipative system is that every jump operator $L_{\n{eff},j}$ with rates resonant at $x_0$ does not correct a unique error state as in standard measurement based stabilizer codes. Instead, it corrects all the error states of weight $x_0$ involving an error on qubit $j$. Another interesting feature is that by fixing the interaction strength $G=G_0$, we also correct error states in the neighborhood of $x_0$. It remains an open question whether one can engineer resonances of more general shapes which are not Lorentzian. This would allow to correct error states of various weights at the same rate and with even less jump operators. Such processes may be achievable using engineered nonlinearities in the shapes of the global laser drives.  

As mentioned above the scale of the Lorentzian is given by $x_0 \kappa_\n{eng}/2\Delta$. We can decrease it, and thus decrease the rate of the undesired process, by increasing the internal states' detuning $\Delta$ or equivalently by increasing the detuning of the oscillators $\delta$ (because of the condition $x_0=\Delta\delta/G^2$).

\medskip
The desired process in Eq.~(6) have a different from than the trickle down solution we proposed in Eqs.~(4). However, they can be shown to be equivalent. It is straightforward to check that the projector $P_{0}^{(j)}$ is related to the projectors $P_{i,k}^{(j)}$ in Eq.~(4) through the relation
\be \label{eq:proj_relation}
    \dyad{1}{0}_i P_{0}^{(j)} = \dyad{1}{1}_i \sum_{k=1}^{K_j} P_{i,k}^{(j)}\,.
\ee
The relation for $P_{1}^{(j)}$ reads identically with the roles of $\ket{0}$ and $\ket{1}$ interchanged. Given that $X_i=\dyad{1}{1}_i+\dyad{0}{0}_i$, the sum of these expressions leads to the intended result. In other words, the desired part of the jump operator in Eq.~(6) realizes to the proposed trickle down solution with the exception that the rates are function of the error weight $j$. To remedy this we can make the global drive $H_2$ multi-tone. Carefully setting the strength $G_x$ and detuning $\delta_x$ of each tone puts several transitions on resonance (according to Eq.~(7)). The optimal number of tones is $\ell$ allowing to address all the desired processes at once. Finally, adding $n$ single-ion addressing laser beams with $n$ auxiliary motional mode $c_i$ (i.e. a separate $H_{3,i}$ from Eq.~\eqref{eq:hamilton3} for every ion) engineers a dissipative process with $n$ jump operators $\big\{L_{\n{eff},i}\big\}_{i=1}^n$. 

\SMsec{Connection to approximate decoders for topological codes}{sec:approx_decoders}
Here, we put our trickle-down solution within the context of the current common ways of performing QEC. As noted earlier, a lookup table with exponentially increasing size is unfeasible even in measurement-feedback QEC. Instead, this paradigm relies on approximate decoding algorithms, which use measurement outcomes to estimate the most likely error subspace into which the system has been projected. In the framework of stabilizer codes, these approximate decoders take as input the binary vector $\vec{s}_i$ (the syndrome) obtained from stabilizer measurements and output a correction unitary $\tilde{U}_i^\dagger$ that ideally approximates the desired $U_i^\dagger$. To further simplify the problem, decoders often incorporate a model of the error process in the system, for example, the locality of errors determined by qubit connectivity and single- and two-qubit error probabilities obtained from device calibration. For codes where each independent error produces a syndrome $\vec{s}_i$ with at most two flagged bits $\vec{s}_i[a]=1=\vec{s}_i[b]$ (as in surface codes), the estimation can be reformulated as a graph problem~\cite{higgott_sparse_2025}. Here, the ${n-k}$ nodes correspond to the entries of $\vec{s}_i$ and the edges represent the errors (or, equivalently, the required corrections). We show an example of such a decoder graph in Fig.~\ref{fig5}.  The decoder's task is to match each pair of flagged vertices with a path (i.e. an error process) with the lowest possible total weight (with the path weight calculated efficiently using the available information). Various algorithms can address this problem, among which the minimum-weight perfect matching algorithm is the most common~\cite{edmonds_maximum_1965,edmonds_paths_1965,dennis_topological_2002,fowler_towards_2012}. For further details about decoders, see Refs.~\cite{iolius_decoding_2024,higgott_sparse_2025}. Translating approximate decoders from measurement-feedback to dissipative QEC, one may define the correction jump operators as 
\be \label{eq:jump_approx_decoder_naive}
    L_{\n{c},i}=\sqrt{\Gqec}\,\tilde{U}_i^\dagger\!(\vec{s}_i)\,P_i
\ee
where $\tilde{U}_i^\dagger\!(\vec{s}_i)$ is the recovery unitary produced by the approximate decoder for the syndrome $\vec{s}_i$ and $P_i$ is the projector onto the $i$th error subspace. These operators are an approximate version of the lookup-table operators in Eq.~(1) with the advantage of limiting the exponential growth of the number of correction jumps with $n$ by excluding the correction of the most unlikely error processes given the error model. However, they still suffer from the requirement $\Gqec\sim\cl{O}(\n{poly}(n))$ (see Eq.~(8)). To improve this aspect, one must consider the topological properties of certain codes. For example, in surface codes, a closed path in the decoder graph may go undetected because no syndrome bit $\vec{s}_i[j]$ is flagged; consequently, such a path would not be recognized as an error. Therefore, to correct a pair of flagged vertices, multiple closing-loop correction paths may be constructed. Conversely, a given unitary $\tilde{U}_i^\dagger\!(\vec{s}_i)$ can recover the logical information from several error subspaces simultaneously, with projectors $\{P_{i,j}\}$ (see examples in Fig.~\ref{fig5}). Thus, the action of approximate decoders in measurement-feedback QEC should be modeled in the dissipative framework using the correction jump operators 
\be \label{eq:jump_approx_decoder}
    L_{\n{c},i}=\sqrt{\Gqec}\,\tilde{U}_i^\dagger\!(\vec{s}_i)\,\sum_j\,P_{i,j}.
\ee
These operators are similar to those proposed in Eq.~(4) with the difference that the $P_{i,j}$ are not necessarily projectors onto subspaces corresponding to a specific error order. Nevertheless, they lead to improved correction performance, analogous to the trickle down solution. That being said, both sets of jump operators in Eqs.~\eqref{eq:jump_approx_decoder_naive} and~\eqref{eq:jump_approx_decoder} are challenging to implement in practice, as they require not only knowledge of the syndrome $\vec{s}_i$ but also the ability to engineer dynamically changing recovery unitaries $\tilde{U}_i^\dagger\!(\vec{s}_i)$. Nonetheless, this discussion highlights two key points: (1) approximate decoders overcome the issue of an exponentially increasing number of correction operators as the code size grows; and (2) a code's topology can induce effects similar to a trickle down correction. These observations also prompt questions regarding measurement-feedback QEC: Is there a sequence of stabilizer measurements whose outcomes can indicate which operation would most effectively reduce the error weight? Can one devise an approximate decoder that functions as a trickle down correction, and would this be more time-efficient? Recent works have begun to address the second question~\cite{jones_improved_2024,ott_decision-tree_2025,beni_tesseract_2025}. For instance, Ref.~\cite{jones_improved_2024} demonstrated that combining outputs from multiple decoders and selecting the best local correction from each can improve both the error suppression factor $\Lambda$ as well as the decoding time. This approach effectively selects the lowest-weight paths across decoder outputs and can be interpreted as minimizing the area of loops in the code (even though not considered as errors). Decoders proposed in Refs.~\cite{ott_decision-tree_2025,beni_tesseract_2025} follow a different strategy: they construct candidate error sequences by adding faults incrementally, guided by a decision tree, until the resulting syndrome matches the observed one. These algorithms could be adapted to prioritize corrections that reduce the overall error weight, rather than reconstructing the full error, thus bringing them closer in spirit to a trickle down correction strategy.

\begin{SCfigure}[50][t!]
    \includegraphics{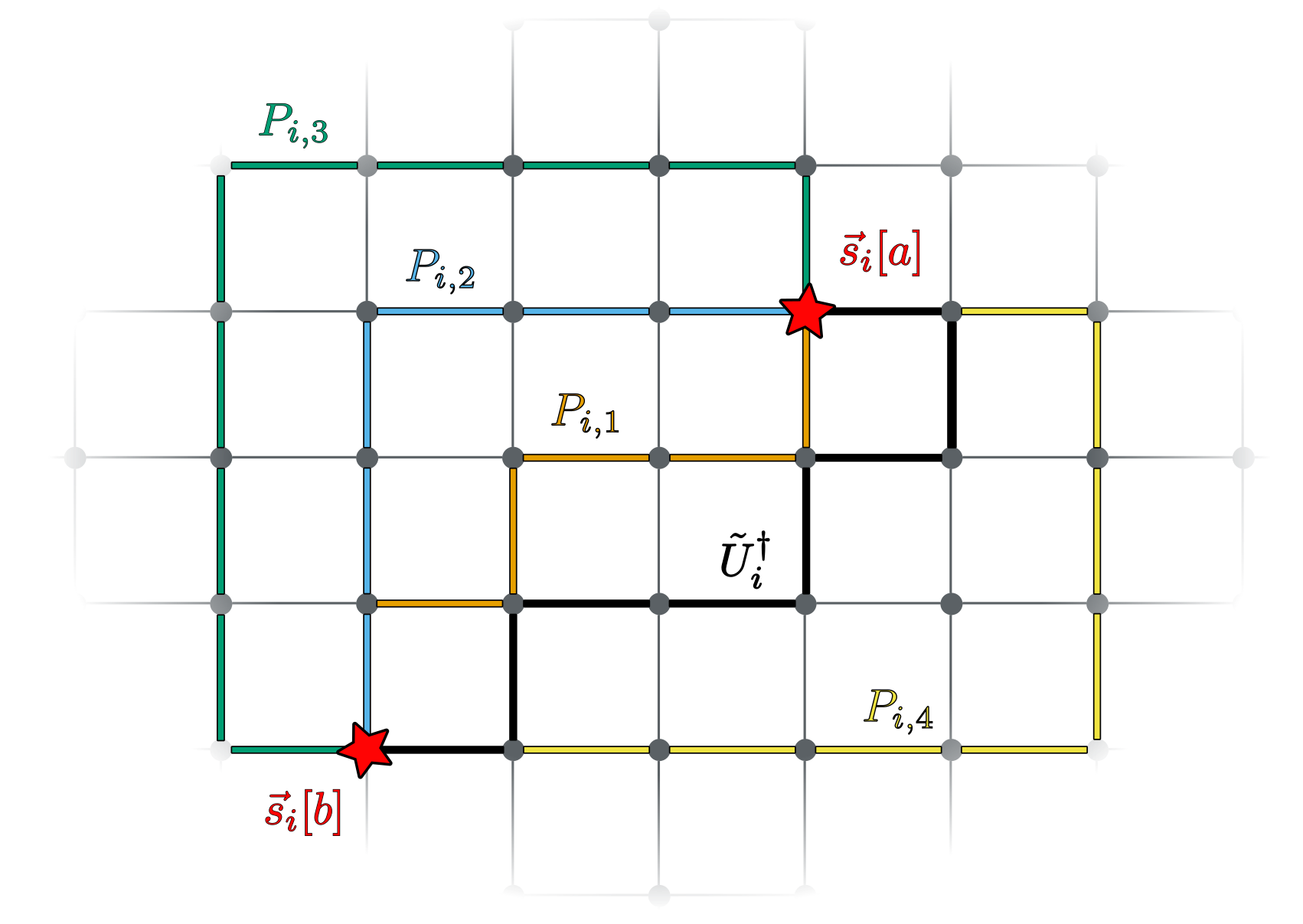}
    \caption{ \textit{Approximate decoders.} Example of a decoder graph for a surface code. Nodes represent syndrome bits $\vec{s}_i$, and edges represent possible error paths (or equivalently, corrections). The decoder's goal is to connect each pair of flagged vertices (i.e., $\vec{s}_i = 1$, marked with stars) via paths of minimal total weight, ideally matching the actual error. The original error is omitted. The decoder outputs a correction path $\tilde{U}_i^\dagger$ for each pair. Due to the topological nature of surface codes, any path differing by a stabilizer (e.g., $P_{i,1}$, $P_{i,2}$, $P_{i,3}$, $P_{i,4}$) is also a valid correction, as such loops are undetectable via stabilizer measurements and do not affect the logical state.}
    \label{fig5}
\end{SCfigure}

\SMsec{Connection to bosonic QEC}{sec:bosonic_qec} 
As mentioned in the main text, the trickle-down QEC is analogous to a cooling process, wherein the system is gradually driven towards its lowest energy configuration. In our context, the system is cooled into a degenerate subspace in a coherent way. The correction of bosonic codes is another type of QEC which is often framed as a type of cooling. This becomes evident through the subsystem decomposition approach, which partitions the infinite-dimensional Hilbert space of a quantum harmonic oscillator into a logical qubit subspace and an auxiliary gauge mode~\cite{pantaleoni_modular_2020}. Rewriting the system dynamics in this form allows one to isolate and analyze the effects of both noise and correction on each subsystem.

A canonical example is the dissipative stabilization of the Cat qubit~\cite{mirrahimi_dynamically_2014}. Consider a quantum harmonic oscillator with annihilation and creation operators $a$ and $a^\dagger$. A dissipative dynamics governed by the jump operator $L_\n{cat}=\sqrt{\Gqec} (a^2-\alpha^2)$ drives the system into a degenerate ground state manifold spanned by the coherent states $\ket{\pm\alpha}$, which form the logical-Z basis of the Cat qubit~\cite{gilles_generation_1994,mirrahimi_dynamically_2014}. In the subsystem decomposition of this code (known as the shifted Fock basis) the logical states can be written as $\ket{\pm\alpha}=\ket{0/1}\otimes\ket{0}$, while for sufficiently large $\alpha$, the jump operator becomes $L_\n{cat}\propto I \otimes b$, where $I$ acts on the logical subspace and $b$ lowers the gauge mode by one excitation~\cite{chamberland_building_2022}. Importantly, in this picture, noise such as dephasing also acts predominantly on the gauge mode, effectively exciting it. That is, $a^\dagger a\approx I\otimes(b^\dagger b + \abs{\alpha}^2 +  \alpha b + \alpha^* b^\dagger)$. Thus, dissipative stabilization of the Cat qubit realizes a trickle-down correction of dephasing noise by gradually removing single excitations from the gauge mode, independent of the gauge mode's state. In contrast, a lookup-table correction approach would require jump operators that measure the gauge mode's excitation level and apply corrections accordingly. The number of correction jumps would then scale with the number of gauge mode excitations which one would like to correct.  

Other bosonic codes also exhibit trickle-down correction. A notable example is the nonlinear generalization of Cat qubits realized through nonlinear reservoir engineering~\cite{stabilization_rojkov_2024}. In their subsystem decomposition, the stabilization operator takes the form $I \otimes f(b^\dagger b) b$, where $f$ is a nonlinear function of the gauge mode's excitation number. Similar to our trapped-ion scheme in Eq.~\eqref{eq:eff_jump_op_proj}, the function $f$, which is tunable via system parameters, controls the rate of trickle-down at different error weights, and has been shown useful to enhance robustness against noise~\cite{stabilization_rojkov_2024,rousseau_enhancing_2025}. 

Our final examples are GKP codes~\cite{gottesman_encoding_2001}. Designed to protect quantum information against small displacements in position and momentum, GKP codes use two dissipative jump operators—each targeting one quadrature~\cite{de_neeve_error_2022,royer_stabilization_2020}. These operators induce a trickle-down correction process, as shown in Ref.~\cite{sivak_real-time_2023}, via a decomposition of the oscillator Hilbert space into error subspaces of increasing boson number gain or loss. Each jump operator acts to transfer the quantum state down this hierarchy, closer to the code subspace. The authors illustrate this process by showing trajectories of various error states through these subspaces during stabilization. The trickle down aspect of these jump operators can also be shown through their subsystem decomposition (known as the shifted grid or Zak basis) which parametrizes the gauge mode in terms of the modular version of position and momentum operators~\cite{pantaleoni_modular_2020,shaw_stabilizer_2024}. Here, as in the Cat code example, the stabilization gradually displaces the gauge mode state towards its ground state. 

In summary, many bosonic QEC schemes developed to date share a trickle-down character. This is largely due to the use of an auxiliary two-level system to mediate correction which intrinsically can only remove one bit of entropy at a time~\cite{de_neeve_modular_2025}. Trickling errors down represents a solution that optimizes the number of error states corrected simultaneously through one entropy removal round. \citet{sivak_real-time_2023} qualified it as being an approach that reduces the control overhead in bosonic codes. In our work, we showed that this reduction is in fact exponential and that it is accompanied by an improvement in the error correction performance. Thus, beyond the well-known benefit of reducing the number of physical systems required to encode a logical qubit, the rapid success of bosonic codes may also stem from the intrinsic efficiency of their QEC protocols. Nonetheless, the auxiliary qubit introduces its own challenges: errors in the qubit can propagate to the oscillator and cause logical failures, thereby undermining fault tolerance~\cite{campagne-ibarcq_quantum_2020,royer_stabilization_2020,sivak_real-time_2023}. Solutions to this issue are a topic of ongoing research~\cite{siegele_robust_2023}.

\SMsec{Future directions}{sec:future_directions}

In this section, we outline several promising avenues for future research inspired by our findings. An interesting prospect is to learn as to whether measurement-feedback QEC can be implemented in a trickle-down fashion. Would this require measurements that differ from the stabilizer ones? If not, is there a classical decoder capable of outputting correction operations that reduce error weight and suitable for real-time decoding of quantum codes?

Our work focused on quantum codes functioning as quantum memories. Although these can be applied to quantum sensing and metrology~\cite{reiter_dissipative_2017}, an intriguing direction for future research is to investigate the impact of trickle-down correction on logical quantum gates for quantum computation. \citet{lebreuilly_autonomous_2021} proved the existence of generalized error-transparent Hamiltonians, showing that if a physical operation transforms the codespace and all relevant error subspaces in the same way, then the correct logical gate is realized upon recovery. In other words, if the operation coherently increases the system's error weight (while retaining it below the uncorrectable level) then dissipative QEC can restore the system such that the overall process implements the intended logical gate. One example of this principle appears in error-corrected gates between GKP codes. These two-qubit gates defined for ideal infinite-energy codewords generate unwanted entanglement when applied to realistic finite-energy GKP states~\cite{gottesman_encoding_2001}. However, dissipative trickle-down QEC has shown to correct much of this spurious entanglement, thereby restoring the logical operation to high-fidelity~\cite{rojkov_two-qubit_2024}. An open question is how such error-transparent operations manifest in other codes and whether they could simplify the experimental requirements for implementing fault-tolerant logical gates.

}{}

\end{document}